\documentclass[graybox, envcountchap]{svmult}

\usepackage{mathptmx}        
\usepackage{amsmath}
\usepackage{amssymb}
\usepackage{color}
\usepackage{helvet}          
\usepackage{courier}         
\usepackage{dirtree}
\usepackage{makeidx}        
\usepackage{graphicx}        
\usepackage{subfig}

\usepackage{multicol}        
\usepackage[bottom]{footmisc}

\usepackage{hyperref}        

\usepackage{alltt} 

\hypersetup{colorlinks=true,urlcolor=blue}

\usepackage[misc]{ifsym}

\makeindex             

\newcommand{\hiMpc}{\,h^{-1}\rm Mpc}
\newcommand{\hiGpc}{\,h^{-1}\rm Gpc}
\newcommand{\Msun}{\, \rm M_\odot}
\newcommand{\hiMsun}{\, h^{-1}\rm M_\odot}
\newcommand{\Mvir}{M_{\rm vir}}

\newcommand{\Hloc}{H_0^{\rm loc}}
\newcommand{\dHloc}{\Delta H_0^{\rm loc}}

\newcommand{\zmax}{z_{\rm max}}

\newcommand{\kmsMpc}{\,{\rm km\ s^{-1}Mpc^{-1}}}

\usepackage[dvipsnames]{xcolor} 

\usepackage{aas_macros}

\begin{document}


\title{Not empty enough: a local void cannot solve the $H_0$ tension}
\author{Dragan Huterer and Hao-Yi Wu}
\institute{Dragan Huterer (\Letter) \at Department of Physics and Leinweber Center for Theoretical Physics, University of Michigan, 450 Church St, Ann Arbor, MI 48103 \email{huterer@umich.edu}
\and Hao-Yi Wu (\Letter) \at Department of Physics, Boise State University, Boise, ID 83725, USA\\ \email{hywu@boisestate.edu}}
%
%
\maketitle

\abstract{
We review arguably the simplest solution for the Hubble tension --- the possibility that we live in a void.  In this scenario, the local Hubble constant $H_0$ is higher than the global value, thus potentially explaining why $H_0$  measured locally by the  distance ladder including Type Ia supernovae (SNIa) would be larger than the value inferred from the cosmic microwave background and other cosmological probes.  In addition, since the local supernova sample is sparse and highly inhomogeneous, the error bars in the local Hubble constant might be larger than previously estimated.  These two effects ---  local matter density and sample inhomogeneity --- constitute the sample variance (or the cosmic variance) of the local Hubble constant measurements.  To investigate these effects explicitly, we have mocked up SNIa observations by exactly matching their actual spatial distribution in a large N-body simulation. We have then investigated whether the sample variance is large enough to explain the Hubble tension.  The answer is resoundingly negative: the typical local variation in $H_0$ is far smaller than what would be required to explain the Hubble tension; the latter would require a 20-$\sigma$ deviation from the expected sample variance.  Equivalently, the void required to explain the Hubble tension would need to be so empty ($\delta\approx-0.8$ on a scale 120 $\hiMpc$) that it would be incompatible with the large-scale structure in a $\Lambda$CDM universe.  Therefore, the possibility that we live in a void does not come close to explaining the Hubble tension.
}



\section{Introduction}


In the currently favored flat $\Lambda$CDM cosmological model, the local measurement of the Hubble constant $H_0$ from the distance ladder \cite{Freedman12, Riess16, Riess:2021jrx} and the global measurement from the CMB anisotropies \cite{WMAP9, Planck13Cosmo, Planck15Cosmo, Planck:2018vyg} should agree. The fact that they do not may well be a harbinger of new physics --- or, at the very least, unexpectedly large systematic errors in either of the two principal measurements.  

Hubble tension has sparked huge interest in cosmology, and a dizzying number of explanations for it have been put forth (see Ref.~\cite{DiValentino:2021izs}, which refers to more than 1000 papers in this regard). However, no single explanation has yet proven to be compelling.  Most of the theoretical explanations try to change the global CMB $H_0$ fit from $\sim 67\kmsMpc$ to $\sim 73\kmsMpc$, the value favored by the distance ladder with SNIa. While doing so,   these new models have to preserve the excellent fit to CMB, baryon acoustic oscillations, and other cosmological data which --- in an unmodified, standard cosmological model --- favor $H_0\sim 67\kmsMpc$. Even in reasonably successful models, it has proven difficult to raise the Hubble constant all the way to $\sim 73\kmsMpc$ (rather than halfway there, to $\sim 70\kmsMpc$) while at the same time improving the goodness of fit to the data to justify the new parameter(s) of the models in question.

In this Chapter, we review arguably the simplest explanation for the Hubble tension --- the possibility that we live in a void empty enough to resolve the Hubble tension. The logic of this explanation goes as follows. There are underdensities and overdensities in the universe. If we live in an underdensity, the local Hubble constant, $\Hloc$, will be higher than the global Hubble constant, as less mass in the void implies a higher expansion rate. This would imply a possible mismatch between $\Hloc$ and the global, ``true'' $H_0$. In addition, given that the local SNIa sample is highly sparse and inhomogeneous in the 3D spatial distribution, the error bar on the local $H_0$ might be larger than previously thought.  The two effects --- local matter density and sample sparsity --- constitute the \textbf{sample variance} (or cosmic variance) in the local $H_0$ measurements.

Our goal is to precisely quantify the sample variance, and we do so in Sec.~\ref{sec:sim}. First, however, we review the physics behind the void explanation in Sec.~\ref{sec:theory}, and briefly review the observational evidence for a void in Sec.~\ref{sec:obs}. We conclude in Sec.~\ref{sec:concl}.

\section{The impact of underdensity on the Hubble constant: theoretical considerations}
\label{sec:theory}

The ``Hubble bubble'' hypothesis refers to the possibility that Milky Way resides in an underdense region in the universe, and the underdensity makes the local expansion rate significantly different from the global expansion rate. If we live in such a void, nearby galaxies would preferentially have positive peculiar velocities (moving away from us faster than expected from the global expansion rate).  This effect can cause the local Hubble constant to be higher than the global value.


The effect of the local density of the universe on the locally measured Hubble constant can be quantified as follows. In an underdense region, the expansion rate is higher than the mean (and vice versa in an overdense region) \cite{Turner92, Wang98, Shi98, CoorayCaldwell06, HuiGreene06, Martinez-Vaquero09, Sinclair10, Courtois13, BenDayan14, Fleury17}. Namely, the relation between the local $H_0$ shift,
$\Delta H_0$, and the local density contrast, $\delta$, is
\begin{equation}
\frac{\Delta H_0}{H_0} =-\frac{1}{3} \delta f(\Omega_M) \Theta(\delta, \Omega_M),
\label{eq:dHloc}
\end{equation}
where $f(\Omega_M)\simeq \Omega_M(a)^{0.55}$ is the growth rate of density perturbations (see Ref.~\cite{Huterer:2022dds} for a review), and $\Theta=1-O(\delta)$ is a non-linear correction which is small for underdensities of typical size. Therefore, an underdensity with $\delta<0$ will automatically lead to a higher measured local expansion rate, that is, $\Delta H_0>0$.  The effect of an under/overdensity described here is qualitatively \textit{guaranteed}, as the local density alone leads to location-dependent local measurements of $H_0$.  This makes the Hubble bubble an appealing model to potentially explain why local measurements of the Hubble constant disagree with the global measurement given by {\em Planck}.

Is it possible that sample variance provides a sufficiently large effect to explain the Hubble tension?  Over the past decade or so, the answer has been arrived at with increasingly realistic numerical and analytical estimates.  The effect of sample variance on the local measurements of $H_0$ was pioneered by Refs.~\cite{Marra13, Wojtak14, Odderskov14} using numerical simulations.  These analyses have strongly hinted that the underdensity required to explain the Hubble tension is extremely unlikely.  In parallel, theorists have investigated the Hubble bubble scenario, relying on simplified analytical models of overdensity.  For example, some work adopted the top-hat spherical model \cite{ZhaiPercival22} or the Lemaitre--Tolman--Bondi (LTB) metric, a variant of the FLRW metric with a spherically symmetric underdensity or overdensity \cite{Marra22, Camarena22,  Castello22}\footnote{Note that some studies combine theoretical and numerical tools \cite{Kenworthy:2019qwq, ZhaiPercival23}.}.  All of these results indicate that it is extremely unlikely that we live in a void that is sufficiently large in size and devoid of matter to explain the Hubble tension. 

However, most of these past analyses of sample variance --- both numerical and theoretical --- have modeled the presence of spherical underdensity but have not attempted to mimic the local $H_0$ measurements particularly closely. Specifically, these studies have been assuming spherical symmetry in the SNIa measurements. However, merely visualizing the highly inhomogeneous distribution of SNIa in the sky brings into question the accuracy of the homogeneity assumption that underlies most past numerical and theoretical treatments.  For example, SNIa distribution is denser in some regions due to historically available observations (e.g.\ Sloan Digital Sky Survey's Stripe 82; see also, for example, Fig.~1 in Ref.~\cite{Soltis:2019ryf}).

Before discussing our work addressing sample inhomogeneity, we briefly review why a direct measurement of the Hubble bubble is challenging.

\section{Do we live in a Hubble bubble? Observations of the local density contrast}
\label{sec:obs}

As far as the observational evidence for the local density contrast is concerned, the following quote by Arthur Eddington
\begin{alltt}\normalfont  
\begin{center}
\textit{Never trust an experimental result until it has been confirmed by theory}
\end{center}
\end{alltt}
\vspace{-14pt}
is perhaps very relevant here. This is because observational evidence for the Hubble bubble must necessarily rely on the estimate of mass density, which is plagued by many observational uncertainties.  The principal challenge is twofold: to convert from light in some electromagnetic band (visible, X-ray, radio, etc.) to mass, and then to map out the three-dimensional distribution of the mass (rather than just the 2D projection on the sky). Despite the massive efforts by observers and modelers, we are still not at the level of accuracy where such direct measurements can be made to make robust claims about the Hubble bubble.


Some observations do hint that the Milky Way might be in an underdense region in the universe, but the results have been  controversial \cite{Zehavi98, Frith03, Jha07, Keenan13, WhitbournShanks14, Carrick15, Bohringer20, WongShanks22}.  Several observations use large-area surveys of galaxy number density to map the radial distribution of mass centered around our location in the universe.  
For example, Ref.~\cite{Keenan13} uses the 600 deg$^2$ galaxy catalog from UKIDSS-LAS, finding a significant local underdensity $\delta=-0.3$ within 200 Mpc.
%
%
Ref.~\cite{Carrick15} uses the full-sky 2M++ catalog and finds no evidence for the local underdensity out to 120 $\hiMpc$.
Ref.~\cite{Kenworthy:2019qwq} uses the distance--redshift relation derived from $\sim1000$ SNe from Pantheon, Foundation, and Carnegie Supernova Project and derived that $|\delta|<27\%$ (5$\sigma$ constraint) on scales larger than 69 $\hiMpc$ ($z=0.023$). In contrast, Ref.~\cite{WongShanks22} find that the local density within $z < 0.075$ is 20\% lower than the mean density of the universe, which would lead to a 3\% correction of the local Hubble constant.

Clearly, on the front of direct observations, things are unsettled, and there is a wide divergence of interpretations of observational measurements. Nevertheless, in the remainder of this review, we demonstrate that, even if we take these results at face value, none of the claimed observed underdensities can alleviate the Hubble tension. As will be described in the next section, resolving the Hubble tension would require an underdensity $\delta \approx -0.8$ within a radius of 120 $\hiMpc$ \cite{Wu:2017fpr}. Such an underdensity is incompatible with any of the observational results and is impossible to achieve within a $\Lambda$CDM framework.  In sum, quantifying the local density contrast is an interesting research subject per se, but it is unlikely to lead to a viable solution for the Hubble tension.

\section{Investigating density contrasts in large-volume N-body simulations}\label{sec:sim}

We investigate the effect of under/overdensities with a controlled experiment, using an N-body simulation of the large-scale structure of the universe that covers many realizations of the supernova sample volume. 

Let us start with some preliminaries. In the presence of a local void, galaxies tend to move away from the center of the void and exhibit positive peculiar velocities.  Let's assume we have $N$ standard candles at comoving distances $r_i$ and radial peculiar velocities  $v_{r, i}$ (i.e.\ radial velocities of galaxies with the global expansion rate of the universe taken out). The shift in the local Hubble constant caused by these peculiar velocities is given by
\begin{equation}
    \Delta\Hloc = \frac{1}{N}\sum_{i=1}^{N} \frac{v_{r,i}}{r_i} ~.
    \label{eq:v_r}
\end{equation}
Clearly, positive net radial peculiar velocities will lead to a positive shift in $H_0$, and vice versa for negative velocities.  Equation (\ref{eq:v_r}) forms a basis for how one can estimate the statistical fluctuations of the observed $H_0$.\footnote{Note that, in SNIa observations, the peculiar velocities of galaxies are usually removed; for example, Ref.~\cite{Riess16} has used the density field reconstructed by the 2M++ survey to correct for the peculiar velocities. Here we discuss the \textit{maximum} impact of the local density by assuming that the peculiar velocities are not corrected for.  Our results thus serve as a conservative upper limit of the sample variance.}

\begin{figure}
\centering
\includegraphics[width=0.8\textwidth]{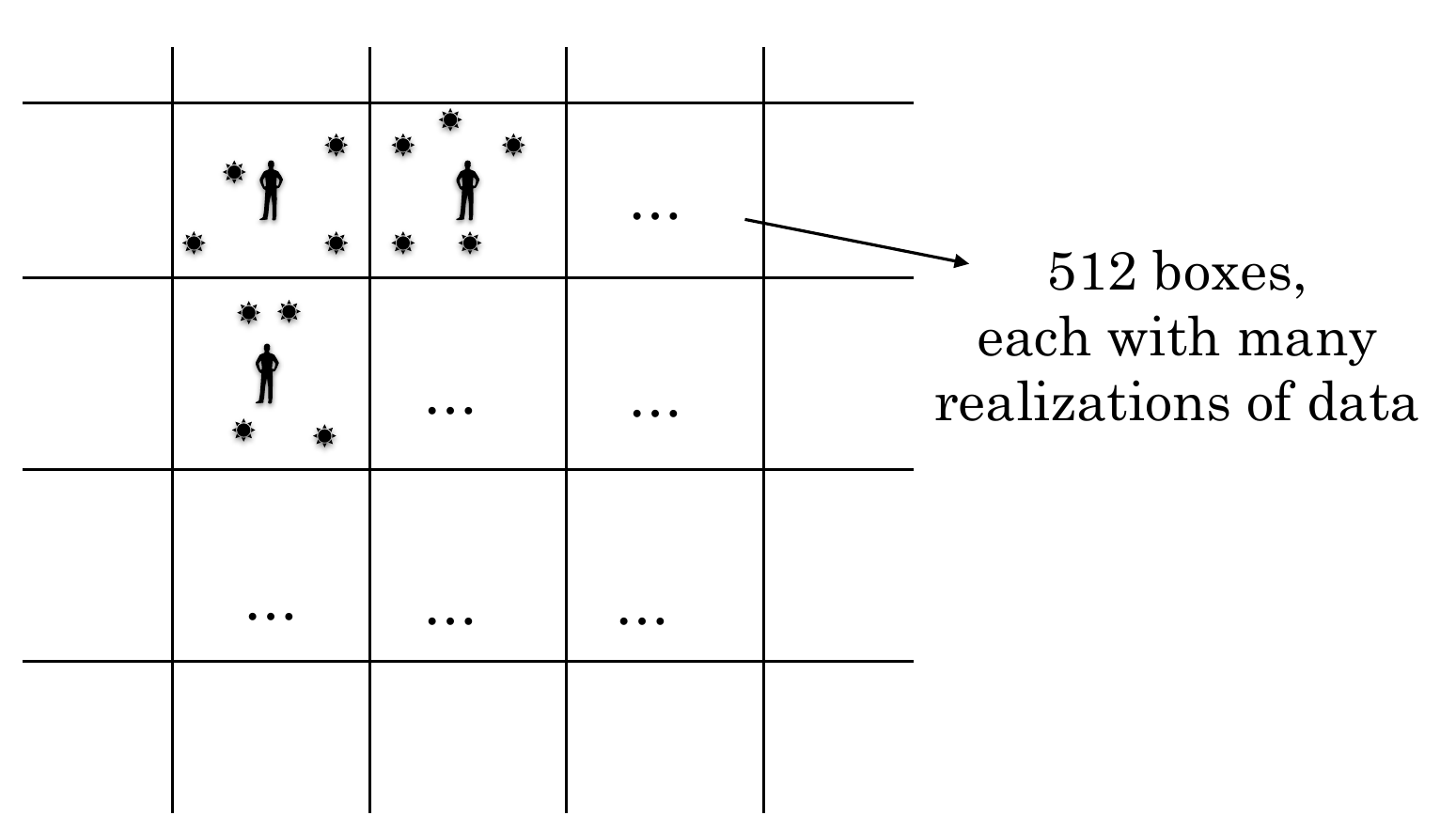}
\caption{Sketch of the simulation procedure in Ref.~\cite{Wu:2017fpr}. We divide the 8 $\hiGpc$ N-body simulation into 512 subvolumes.  In each subvolume, we place an observer at the center, use $\sim$3000 rotations of the global coordinates of the observed SNIa (not shown in this sketch), and find the closest Milky Way-mass halo to each SNIa.  We then use the known properties of that halo --- its distance from the observer and its peculiar velocity --- as proxies for those of SNIa, and use Eq.~\ref{eq:v_r} to estimate the Hubble constant shift.}
\label{fig:dH0_sketch}
\end{figure}

Here we recapitulate the results of Wu \& Huterer \cite{Wu:2017fpr}.  We address two aspects of the sample variance: (1) variance caused by the local density contrast, and (2) variance caused by the inhomogeneous selection of the supernova sample ``Supercal'' \cite{Scolnic14, Scolnic15}, which is used in the Riess et al.~(2016) analysis \cite{Riess16}.  The former effect is irreducible, while the latter effect can be reduced with future larger samples.  

To address the contribution of the local density contrast to the Hubble constant, we perform our study on a large N-body simulation. Our general idea is to place the observer at many independent locations in the simulation, mimic the SNIa observations around each observer, and measure the statistical distribution of the inferred Hubble constant with respect to the global value (i.e.\ in $\Delta H_0$ from Eq.~\ref{eq:dHloc}). The advantage of this approach is that we can mimic the SNIa observations precisely by selecting halos in our simulation that stand for SNIa with their (near-)exact three-dimensional locations around the observer. This enables measurements of $\Delta H_0$ that account for the sample variance of actual SNIa observations very precisely. Given the size of the numerical simulation and our consequent ability to place many independent observed volumes in it, we also have great statistical power at our disposal to quantify the effect of sample variance.


We use the public release of the Dark Sky simulations \cite{Skillman14}.  Specifically, we use the $(8\hiGpc)^3$ volume with $10^{12}$ particles and a mass resolution $3.9\times 10^{10}~\hiMsun$.  The simulation assumes a standard flat $\Lambda$CDM model consistent with {\em Planck} 2016 \cite{Planck1605.02985}.  We divide this $(8\hiGpc)^3$ volume into 512 subvolumes of $(1\hiGpc)^3$.  We then choose the halo with virial mass $\Mvir \in [10^{12.3}, 10^{12.4}]~\Msun$ closest to the center of each subvolume as our observer.  This choice simulates 512 independent observers living in Milky Way-mass halos, each with a separate subvolume of the large-scale structure out to the distance of interest ($\zmax = 0.15$). For the host halos of SNIa, we also use Milky Way-mass halos with $\Mvir \in [10^{12.3}, 10^{12.4}]~\Msun$, and we have explicitly checked that using all halos above $10^{13}~\Msun$ does not change the result.  We then simulate the Supercal supernova observation in each subvolume.  

Our procedure then goes as follows (see also Fig.~\ref{fig:dH0_sketch}):

\begin{enumerate}

\item Divide the 8$\hiGpc$ Dark Sky box into 512 subvolumes as mentioned above; each Milky Way-mass halo in every subvolume is a possible host of an SNIa. Then, for each subvolume:

\item Select a random orientation of the overall SNIa sky coordinates with respect to the simulation's Cartesian coordinates.

\item Assign each SNIa to the closest halo in the subvolume based on its 3D coordinates reported in the Supercal dataset release \cite{Scolnic14, Scolnic15}.

\item Calculate the $\Delta\Hloc$ from the radial velocities of SNIa host halos using the relation
\begin{equation}
    \Delta\Hloc = \frac{1}{N}\sum_{i=1}^{N} \frac{v_{r,i}}{r_i} ~,
\end{equation}
where $r_i$ and $v_{r,i}$ are the comoving distance to, and the peculiar velocity of, the halo that hosts the $i^{th}$ SNIa. We average over the $\Delta\Hloc$ of all supernovae from this particular orientation for this observer.  (The actual analysis includes additional details; see Ref.~\cite{Wu:2017fpr}.)
   
\item Go to step 2., repeat the measurements for many different coordinate-system orientations and obtain the histogram of $\dHloc$ of different orientations from this particular subvolume.

\item Go to step 1., repeat the measurements for the 512 non-overlapping subvolumes and obtain the distribution of $\dHloc$ from all subvolumes and all orientations.

\end{enumerate}

The 512 subvolumes account for the variance in the local density in a $\Lambda$CDM universe, and the $\sim$ 3000 SNIa coordinate-system orientations in each subvolume account for the skewed redshift distribution and the sparse angular distribution of the SNIa sample.  The full procedure therefore corresponds to $\sim$1.5 million simulations of inferring the Hubble constant from local SNIa based on the Supercal sample ($0.023<z<0.15$).

\begin{figure*}[t]
  \centering
  \includegraphics[width=0.9\textwidth]{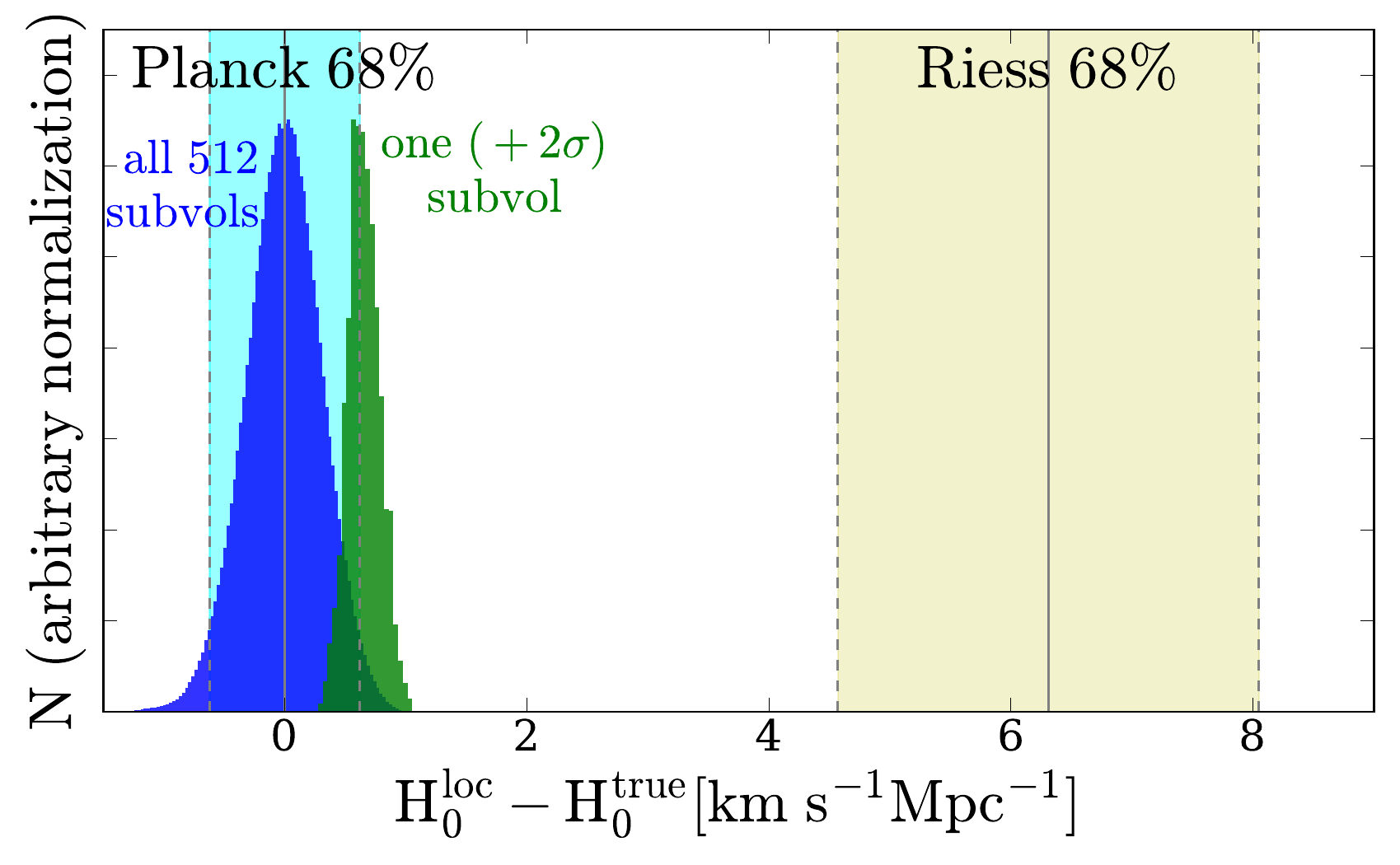}
\caption{Sample variance in $\dHloc$ from the simulations of Wu \& Huterer \cite{Wu:2017fpr}, compared to the {\em Planck} and distance-ladder (\cite{Riess16}, ``Riess'' in the plot) error bars, assuming {\em Planck}'s $H_0$ \cite{Planck15Cosmo} to be the true global value. The blue histogram shows 3240 rotations of the SNIa coordinate system from 512 subvolumes derived from the 8 $\hiGpc$ Dark Sky box, corresponding to $\sim$1.5 million SN-to-halo coordinate-system configurations. The green histogram shows a particularly underdense subvolume with a high $\dHloc$ at the 2-$\sigma$ level relative to the mean of all subvolumes.  The sample variance in $\Hloc$ is much smaller than the difference between SNIa and CMB measurements.  Adopted from Ref.~\cite{Wu:2017fpr}.}
\label{fig:dH0_histograms}
\end{figure*}

Fig.~\ref{fig:dH0_histograms} shows the histogram of the distributions of $\Delta H_0$ for all subvolumes and SNIa coordinate-system orientations.  The overall distribution in the Hubble constant shifts, shown with the dark blue histogram, indicates that the sample variance in the local $\Hloc$ measurements is much smaller than the amount that could explain the Hubble tension (i.e.\ one that would reach out to the Riess et al.\ value, shown as the vertical beige bar).  For example, even a void that is underdense at the $\sim 2\sigma$ level relative to the mean (the green histogram) would lead to the locally measured Hubble constant, $\Hloc$, that is only slightly larger than the mean value, assumed here to be {\em Planck}'s $H_0\simeq 67\kmsMpc$.

We have, in fact, been able to measure the standard deviation of the measured Hubble constant very precisely,
\begin{equation}
  \sigma_{\rm sample\, variance}=0.31\kmsMpc.
\end{equation}
In other words, $|H_0^{\rm Planck}-\Hloc|/\sigma_{\rm sample\,  variance}\simeq 20$. Therefore, \textbf{sample variance is about 20 times too small to explain the Hubble tension}.  Interestingly, using completely different approaches based on perturbation theory, Refs.~\cite{ZhaiPercival22, ZhaiPercival23} find a very similar result: a sample variance of $\sim 0.4 \kmsMpc$ for the Pantheon sample (also see \cite{CamarenaMarra18}).

\begin{figure*}[t]
  \centering
  \includegraphics[width=0.9\textwidth]{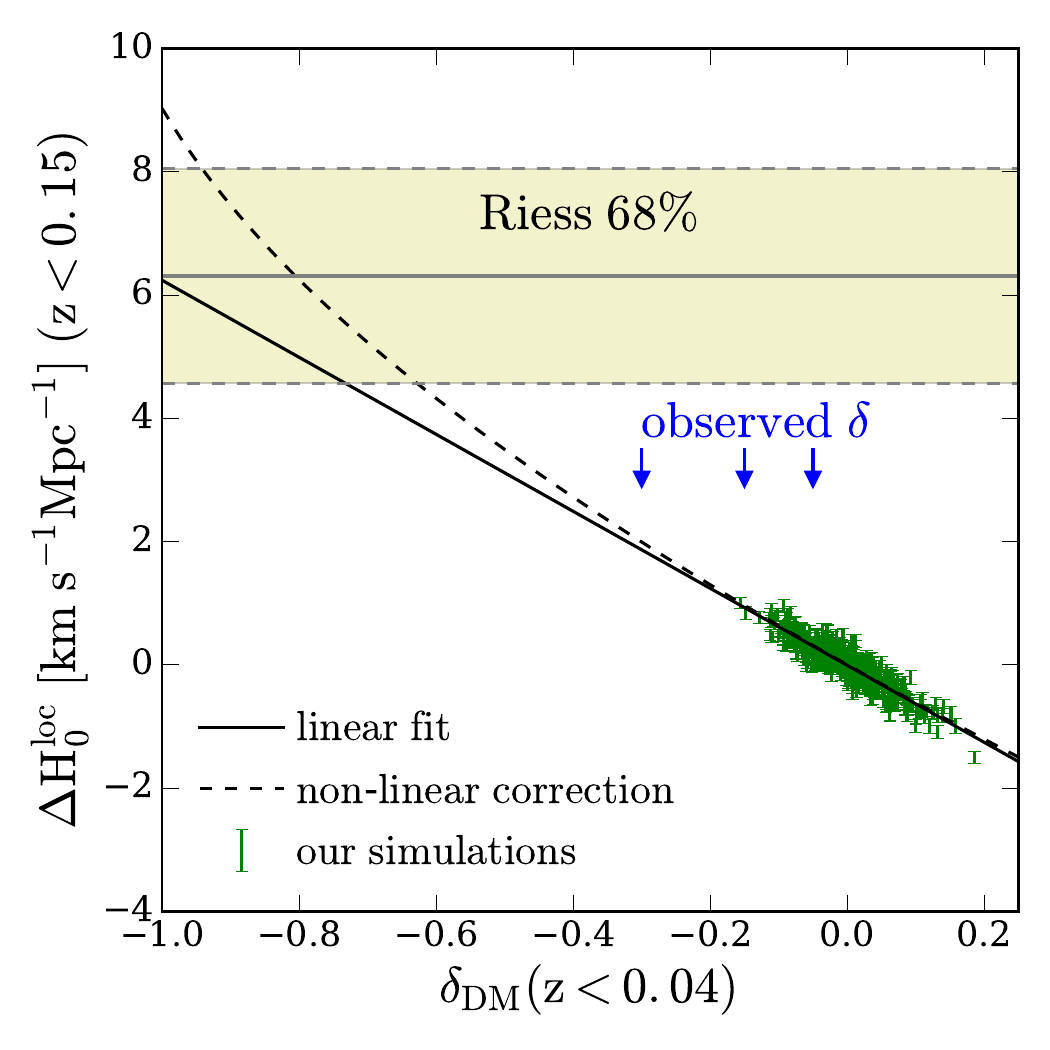}
\caption{Shift in local Hubble constant vs.~local density contrast derived from the mock Supercal sample created from the 8 $\hiGpc$ Dark Sky box (green error bars).  The blue arrows mark the local density inferred from Refs.~\cite{Keenan13, Carrick15, WhitbournShanks14}. The black solid line shows the linear fit, and the dashed curve shows the non-linear correction.  A local Hubble constant shift of $\Delta\Hloc\simeq 6\kmsMpc$ required to explain the Hubble tension would consequently require an underdensity of $\delta\approx-0.8$ with a radius of 120 $\hiMpc$, which is incompatible with the cosmic structure in a $\Lambda$CDM universe.  Adopted from Ref.~\cite{Wu:2017fpr}. }
\label{fig:dH0_delta}
\end{figure*}

How ``empty'' should a local void be to resolve the Hubble tension?   Fig.~\ref{fig:dH0_delta} shows the relation between the change of the local Hubble constant, $\Delta \Hloc$, and the local density contrast, $\delta$.  For each of the 512 subvolumes, we measure the dark matter density contrast at $z<0.04$ ($120 \hiMpc$) and perform mock measurements of $\Hloc$ using the Supercal supernova sample ($z<0.15$).  The small green error bars correspond to the 512 simulation subvolumes, and the black line is the linear fit.  The dashed curve includes the non-linear correction.  This slope is different from that expected from the perturbation theory because we explicitly consider the sample selection, which simple perturbation-theory calculations do not account for.  

The horizontal line shows that to account for the 6 $\kmsMpc$ shift requires an underdensity of $\delta\approx-0.8$ with a radius of 120 $\hiMpc$, which is incompatible with the density fluctuations in the $\Lambda$CDM model.  The blue arrows correspond to observational results from Refs.~\cite{Keenan13, Carrick15, WhitbournShanks14}, and none of them can cause a 6 $\kmsMpc$ shift.

Why does the sample variance of local measurements with SNIa have such a small effect?  The reason is in the fact the Supercal SNIa sample goes out to $z=0.15$ (corresponding to distances of about $\sim 500\hiMpc$). 
Averaged over such a large volume, overdensities and underdensities are not very pronounced, and the variations in the locally determined Hubble constant are correspondingly small.

This result, and the related ones cited above, put a nail in the coffin of the sample-variance explanations for the Hubble tension. Because the sample-variance (or void) explanation was arguably the simplest one, it leads to an exciting situation that the true explanation, whatever it is, is likely to only be more  ``exotic'' and, overall, more unexpected.

\section{Summary}\label{sec:concl}

We have reviewed what is logically the simplest explanation for the Hubble tension --- the notion that we might be simply seeing the effect of sample variance in the distance-ladder-based measurements. Sample variance affects the last rung of the distance ladder, whose distance--redshift relation can be impacted by local density fluctuations. In the sample-variance explanation scenario, 1) we live in an underdense region of the universe, which has an enhanced locally measured Hubble constant, and 2) the local supernova sample is sparse and highly inhomogeneous so that the error bars for the local Hubble constant might be larger than previously estimated.

We have first illustrated the formalism that links the local density to the measured Hubble constant and have reviewed the methodology adopted by the community over the past decade. We have then reviewed recent observational results of the local matter density contrast, which have been inconclusive because it is difficult to infer the three-dimensional distribution of mass density from galaxies.

Proceeding to discuss our own work on the subject \cite{Wu:2017fpr}, we set out our goal:  to use an N-body simulation to mimic the actual SNIa observations used for $H_0$ measurements. In the simulation, we place an observer in many independent subvolumes, each containing the 3D distribution of SNIa that precisely matches that used in actual observations.  This automatically takes into account the effect of over/underdensities, as well as the inhomogeneous distribution of supernovae. Moreover, each SNIa is represented by a host halo, whose peculiar velocity is known in the simulation. Therefore, we have access to a simulated distance-ladder procedure that takes full account of sample variance, including how it mimics actual SNIa observations. The procedure is also statistically powerful ---  we use a total of 1.5 million realizations of SNIa observations; see Sec.~\ref{sec:sim}.

Combining large-volume N-body $\Lambda$CDM simulations with recent SNIa data, we have found that the effect of the sample variance is small. Specifically, the resulting standard deviation in the Hubble constant (the square root of sample variance) is 
\begin{equation}
    \sigma_{\rm sample\, variance} =  0.31 ~\kmsMpc.
\end{equation}  
To explain the 6 $\kmsMpc$ Hubble tension would require a $20\sigma$ effect in sample variance. Sample variance is therefore an extremely unlikely explanation for the Hubble tension. 
    
In addition, we have found that resolving the 6 $\kmsMpc$ Hubble tension requires an underdensity $\delta=-0.8$ out to a radius 120 $\hiMpc$. The underdensity required to resolve the Hubble tension is so low that it is incompatible with the large-scale structure in a $\Lambda$CDM universe. Although the observational evidence for local underdensity is still controversial,
even the most extreme observational claims for the local underdensity could not explain the Hubble tension.  

Ruling out the local void is an exciting development in the nascent field of Hubble tension, as it removes this very simple explanation and a ``guaranteed" effect in cosmology, leaving more exotic or surprising solutions. Future research should focus on the systematics of supernova and CMB measurements, as well as the exploration of new physics beyond the standard model of cosmology.

\begin{acknowledgement}
DH is supported by NASA grant under contract 19-ATP19-0058 and DOE under contract DE-FG02-95ER40899.
HW is supported by the DOE grant under contract DE-SC0021916. 
\end{acknowledgement}

\bibliographystyle{plain}
\bibliography{refs}


\end{document}